\documentclass[prc,twocolumn,superscriptaddress,showpacs,aps,floatfix]{revtex4}
\usepackage{graphicx}% Include figure files
\begin{document}

%\documentclass[preprint,12pt]{elsarticle}
%\usepackage{amssymb}
%\usepackage{graphicx}
%\journal{Nucl. Phys. A}
%\begin{document}
%\begin{frontmatter}

\title{Direct observation of long-lived isomers in $^{212}$Bi}

\author{L.~Chen}
%\email{lchen@comp.tamu.edu}
\affiliation{GSI Helmholtzzentrum f\"ur Schwerionenforschung, Planckstra\ss e 1, 64291 Darmstadt, Germany}
\affiliation{II Physikalisches Institut, Justus-Liebig-Universit\"at Gie\ss en,
%Heinrich-Buff-Ring 16, 
35392 Gie\ss en, Germany}
\affiliation{Cyclotron Institute, Texas A \& M University, Texas 77843, USA}

\author{P.M.~Walker}
\affiliation{Department of Physics, University of Surrey, Guildford, Surrey GU2 7XH, United Kingdom}
%\affiliation{CERN, CH-11263 Geneva 23, Switzerland}

\author{H.~Geissel}
\affiliation{GSI Helmholtzzentrum f\"ur Schwerionenforschung, Planckstra\ss e 1, 64291 Darmstadt, Germany}
\affiliation{II Physikalisches Institut, Justus-Liebig-Universit\"at Gie\ss en,
%Heinrich-Buff-Ring 16, 
35392 Gie\ss en, Germany}

\author{Yu.A.~Litvinov}
\affiliation{GSI Helmholtzzentrum f\"ur Schwerionenforschung, Planckstra\ss e 1, 64291 Darmstadt, Germany}
\affiliation{Max-Planck-Institut f\"ur Kernphysik, Saupfercheckweg 1, 69117 Heidelberg, Germany}

\author{K.~Beckert}
\affiliation{GSI Helmholtzzentrum f\"ur Schwerionenforschung, Planckstra\ss e 1, 64291 Darmstadt, Germany}

\author{P.~Beller}
\affiliation{GSI Helmholtzzentrum f\"ur Schwerionenforschung, Planckstra\ss e 1, 64291 Darmstadt, Germany}

\author{F.~Bosch}
\affiliation{GSI Helmholtzzentrum f\"ur Schwerionenforschung, Planckstra\ss e 1, 64291 Darmstadt, Germany}

\author{D.~Boutin}
\affiliation{GSI Helmholtzzentrum f\"ur Schwerionenforschung, Planckstra\ss e 1, 64291 Darmstadt, Germany}

\author{L.~Caceres}
\affiliation{GSI Helmholtzzentrum f\"ur Schwerionenforschung, Planckstra\ss e 1, 64291 Darmstadt, Germany}

\author{J.J.~Carroll}
\affiliation{US Army Research Laboratory, Adelphi, MD 20783, USA}

\author{D.M.~Cullen}
\affiliation{Schuster Laboratory, University of Manchester, Manchester M13 9PL, United Kingdom}

\author{I.J.~Cullen}
\affiliation{Department of Physics, University of Surrey, Guildford, Surrey GU2 7XH, United Kingdom}

\author{B.~Franzke}
\affiliation{GSI Helmholtzzentrum f\"ur Schwerionenforschung, Planckstra\ss e 1, 64291 Darmstadt, Germany}

\author{J.~Gerl}
\affiliation{GSI Helmholtzzentrum f\"ur Schwerionenforschung, Planckstra\ss e 1, 64291 Darmstadt, Germany}

\author{M.~G\'orska}
\affiliation{GSI Helmholtzzentrum f\"ur Schwerionenforschung, Planckstra\ss e 1, 64291 Darmstadt, Germany}

\author{G.A.~Jones}
\affiliation{Department of Physics, University of Surrey, Guildford, Surrey GU2 7XH, United Kingdom}

\author{A.~Kishada}
\affiliation{Schuster Laboratory, University of Manchester, Manchester M13 9PL, United Kingdom}

\author{R.~Kn\"{o}bel}
\affiliation{GSI Helmholtzzentrum f\"ur Schwerionenforschung, Planckstra\ss e 1, 64291 Darmstadt, Germany}

\author{C.~Kozhuharov}
\affiliation{GSI Helmholtzzentrum f\"ur Schwerionenforschung, Planckstra\ss e 1, 64291 Darmstadt, Germany}

\author{J.~Kurcewicz}
\affiliation{GSI Helmholtzzentrum f\"ur Schwerionenforschung, Planckstra\ss e 1, 64291 Darmstadt, Germany}

\author{S.A.~Litvinov}
\affiliation{GSI Helmholtzzentrum f\"ur Schwerionenforschung, Planckstra\ss e 1, 64291 Darmstadt, Germany}

\author{Z.~Liu}
\affiliation{Department of Physics, University of Surrey, Guildford, Surrey GU2 7XH, United Kingdom}
\affiliation{School of Physics and Astronomy, University of Edinburgh, Edinburgh EH9 3JZ, United Kingdom}

\author{S.~Mandal}
\affiliation{GSI Helmholtzzentrum f\"ur Schwerionenforschung, Planckstra\ss e 1, 64291 Darmstadt, Germany}

\author{F.~Montes}
\affiliation{Michigan State University, East Lansing, Michigan 48824, USA}

\author{G.~M\"unzenberg}
\affiliation{GSI Helmholtzzentrum f\"ur Schwerionenforschung, Planckstra\ss e 1, 64291 Darmstadt, Germany}

\author{F.~Nolden}
\affiliation{GSI Helmholtzzentrum f\"ur Schwerionenforschung, Planckstra\ss e 1, 64291 Darmstadt, Germany}

\author{T.~Ohtsubo}
\affiliation{Department of Physics, Niigata University, Niigata 950-2181, Japan}

\author{Z.~Patyk}
\affiliation{National Centre for Nuclear Research, Ho$\dot{z}$a 69, 00-681 Warszawa, Poland}

\author{W.R.~Pla\ss}
\affiliation{II Physikalisches Institut, Justus-Liebig-Universit\"at Gie\ss en,
%Heinrich-Buff-Ring 16, 
35392 Gie\ss en, Germany}

\author{Zs.~Podoly\'ak}
\affiliation{Department of Physics, University of Surrey, Guildford, Surrey GU2 7XH, United Kingdom}

\author{S.~Rigby}
\affiliation{Schuster Laboratory, University of Manchester, Manchester M13 9PL, United Kingdom}

\author{N.~Saito}
\affiliation{GSI Helmholtzzentrum f\"ur Schwerionenforschung, Planckstra\ss e 1, 64291 Darmstadt, Germany}

\author{T.~Saito}
\affiliation{GSI Helmholtzzentrum f\"ur Schwerionenforschung, Planckstra\ss e 1, 64291 Darmstadt, Germany}

\author{C.~Scheidenberger}
\affiliation{GSI Helmholtzzentrum f\"ur Schwerionenforschung, Planckstra\ss e 1, 64291 Darmstadt, Germany}
\affiliation{II Physikalisches Institut, Justus-Liebig-Universit\"at Gie\ss en,
%Heinrich-Buff-Ring 16, 
35392 Gie\ss en, Germany}

\author{E.C. Simpson}
\affiliation{Department of Physics, University of Surrey, Guildford, Surrey GU2 7XH, United Kingdom}

\author{M.~Shindo}
\affiliation{Department of Physics, University of Tokyo, Tokyo 113-0033, Japan}

\author{M.~Steck}
\affiliation{GSI Helmholtzzentrum f\"ur Schwerionenforschung, Planckstra\ss e 1, 64291 Darmstadt, Germany}

\author{B.~Sun}
\affiliation{GSI Helmholtzzentrum f\"ur Schwerionenforschung, Planckstra\ss e 1, 64291 Darmstadt, Germany}

\author{S.J.~Williams}
\affiliation{Department of Physics, University of Surrey, 
Guildford, Surrey GU2 7XH, United Kingdom}

\author{H.~Weick}
\affiliation{GSI Helmholtzzentrum f\"ur Schwerionenforschung, Planckstra\ss e 1, 64291 Darmstadt, Germany}

\author{M.~Winkler}
\affiliation{GSI Helmholtzzentrum f\"ur Schwerionenforschung, Planckstra\ss e 1, 64291 Darmstadt, Germany}

\author{H.-J.~Wollersheim}
\affiliation{GSI Helmholtzzentrum f\"ur Schwerionenforschung, Planckstra\ss e 1, 64291 Darmstadt, Germany}

\author{T.~Yamaguchi}
\affiliation{Graduate School of Science and Engineering, Saitama University, Saitama 338-8570, Japan}

%\fntext{Present address: Cyclotron Institute, Texas A \& M University, Texas 77843, USA}

\begin{abstract}
Long-lived isomers in $^{212}$Bi have been studied following $^{238}$U projectile fragmentation at 670 MeV per nucleon. The fragmentation products were injected as highly charged ions into the GSI storage ring, giving access to masses and half-lives. While the excitation energy of the first isomer of $^{212}$Bi was confirmed, the second isomer was observed at 1478(30) keV, in contrast to the previously accepted value of $>$1910 keV. It was also found to have an extended Lorentz-corrected in-ring half-life $>$30 min, compared to 7.0(3) min for the neutral atom. Both the energy and half-life differences can be understood as being due a substantial, though previously unrecognised, internal decay branch for neutral atoms. Earlier shell-model calculations are now found to give good agreement with the isomer excitation energy. Furthermore, these and new calculations predict the existence of states at slightly higher energy that could facilitate isomer de-excitation studies.
\end{abstract}

\pacs{21.10.-k, 21.60.-n, 25.70.Mn, 27.70.+w, 29.20.Dh}

\maketitle

%\begin{keyword}21.10.Dr,  21.60.-n, 32.10.Bi
%\end{keyword}
%\end{frontmatter}

Isomers are long-lived excited states of atomic nuclei \cite{Wa99}. Their inhibited decays arise from nuclear shape changes and angular momentum (spin) selection rules, leading to a special role in nuclear physics and  astrophysics research \cite{Ap05} and the the possibility of novel applications such as energy-storage devices, if appropriate conditions can be realised \cite{Wa05}. The understanding of the structure and properties of extreme isomers, combining a long half-life with high spin and/or excitation energy, is a key part of these investigations.

The nuclide $^{212}$Bi has an excited state with a unique combination of properties for a spherical nucleus: $I \geq 16$, and $t_{1/2} = 7.0(3)$ min \cite{Br05}. Nevertheless, it remains poorly characterised, with, for example, unmeasured excitation energy. This quantity is needed foremost to test the predictive power of nuclear shell-model calculations \cite{Wa91}, which themselves are required for modeling elemental synthesis in explosive rapid-neutron-capture ($r$-process) astrophysical environments \cite{Ar07,Go12}. With only four nucleons (one proton and three neutrons) outside the doubly magic core of $^{208}$Pb, shell-model calculations should be reliable for $^{212}$Bi. Surprisingly, however, the estimate of $E^\ast >$1910 from its $\beta$-decay rate \cite{Br05} is substantially different from the calculated energy of 1496 keV for the best isomer candidate, with $I^\pi = 18^-$ \cite{Wa91}. Notwithstanding these contrary indications, if the differences could be properly understood, then the possibility of exploiting the isomer for energy-release studies could be addressed.

In the quest for the manipulation of nuclear isomers with low-energy electromagnetic probes, isomer targets had seemed the most promising, for example $^{180m}$Ta, with $t_{1/2} > 7\times 10^{15}$ yr \cite{Hu06}, and $^{178m2}$Hf, with $t_{1/2} = 31$ yr \cite{Au03}. However, the former requires $>$1 MeV photons \cite{Be99}, and initial claims for the latter \cite{Co99} have been refuted \cite{Ah05,Ca09}. Nevertheless, with new radioactive-beam developments, the half-life requirement is now less stringent. This has been demonstrated through the induced depopulation of $^{68m}$Cu, with $t_{1/2} = 3.8$ min \cite{St07}, exploiting the Coulomb field of virtual photons. Half-lives $>$1 min and high spin appear to be advantageous for isomer separation and detection. If the isomer de-excitation pathway involves $\gamma$-ray emission, then high spin favors multiple $\gamma$ rays, leading to improved detection capabilities. Furthermore, induced isomer depopulation requires the existence of structurally related states of slightly higher energy that can be excited through low-multipole transitions. This mitigates against isomers in deformed nuclei, where the $K$ quantum number (the spin projection on the symmetry axis) limits such possibilities. In contrast, in spherical nuclei the seniority scheme of angular momentum coupling can lead to the desired situation, especially in odd-$A$ and odd-odd nuclei. A notable case is $^{93m}$Mo where an $I^\pi = 17/2^+$ state is just 5 keV higher in energy than a 21/2$^+$ isomer, and nuclear excitation by electron capture (NEEC) could play a decisive role \cite{Go07,Pa07,Ha11,Ka12}. In this context, the extreme properties ($I \geq 16$, $t_{1/2} = 7.0$ min) of the $^{212}$Bi isomer are of special interest. 

The direct observation of highly charged stored ions has been shown to be a powerful experimental technique for isomer studies \cite{Re10}. Individual ions can be identified even without decay events, mass measurements provide isomer excitation energies with no reliance on other level-scheme information, and half-lives can be determined. This opportunity, uniquely available at the experimental storage ring (ESR) at GSI, formed the basis for the present study.
A 670 MeV per nucleon $^{238}$U beam was
extracted from the heavy-ion synchrotron SIS \cite{SIS}, with a
maximum intensity of $2\times$10$^9$ ions per spill, and focused on a 4
g/cm$^2$ beryllium production target placed at the entrance of the
fragment separator FRS \cite{FRS}. Fast extraction was used with a spill
length of 300~ns and a typical repetition rate of 0.25 per minute.
The fragments of interest were separated in flight with the FRS
and injected into the 108-m circumference ESR \cite{ESR}.
Two-fold magnetic-rigidity analysis in the FRS, combined with 
energy loss in a 50 mg/cm$^2$ plastic degrader, restricted the transmitted
range of elements to those between gold and uranium. Some of the
measurements were also performed with pure magnetic-rigidity
separation of the FRS. The main goal of this experiment
was to perform mass measurements \cite{Ch12}. Therefore, the separation
conditions at the FRS were selected such that, in addition to the
nuclei of interest, a sufficient number of nuclides with well-known masses
were injected into the ESR and recorded in the same revolution-frequency
spectra.  

Electron cooling was applied to the ion beam stored in the ESR.
This forces the circulating ions to the same mean velocity, which
is determined by the terminal voltage of the electron
cooler. In the present experiment, the ions had a velocity
about 70\% that of light, corresponding to kinetic energies in the 
range of 360 to 400 MeV per nucleon, and an orbital period close 
to 0.5 $\mu$s. The electron-cooled ions had an equilibrium 
velocity spread ($\Delta v/v$) of approximately 5$\times10^{-7}$.

Each peak in the revolution-frequency spectrum corresponds to 
a specific mass-to-charge ratio. For the measurement of frequency spectra, the
current signals induced at each revolution by the circulating few-electron 
heavy ions were recorded on two metallic pick-up plates in dipolar
arrangement (Schottky pick-ups). The signal from the pick-up plates was tuned in 
a resonance circuit, amplified, summed and shifted down by a frequency of
about 59~MHz. The resulting signal was split into two parts, one for on-line
monitoring and one for off-line analysis. 
Additional experimental details about this technique of Schottky mass spectrometry 
are given in Refs.~\cite{Ch12,Li05,Li11}, including a description of the mass
calibration procedure.
Associated discoveries of neutron-rich isotopes \cite{Ch10} and mass measurements \cite{Ch12}, 
together with an isomer in $^{213}$Bi, came from the same data set.

Two isomers in $^{212}$Bi are known from the $\alpha$- and $\beta$-decay studies of Baisden et al.~\cite{Ba78}, confirmed by Eskola et al.~\cite{Es84}. The most recent evaluation is that of Browne \cite{Br05}. The first isomer, with $t_{1/2} = 25.0(2)$ min, has a tentative $I^\pi = (8^-,9^-)$ assignment and an excitation energy of 250(30) keV \cite{Au03}. The present storage-ring measurement of 239(30) keV is consistent with the known value. The new data are illustrated in Figs.~1 and 2. Fig.~1 shows the time dependence of the Schottky frequency signal. 
The cooling time depends strongly on the matching between the velocity of the injected ions and the velocity of the cooler electrons \cite{Po90}.
The trace labeled $^{212m2}$Bi$^{81+}$ is from a single ion which was quickly cooled and persisted for the whole observation time, whereas the $^{212m1}$Bi and $^{212g}$Bi ions took longer to cool. Then, after $\sim$140 s, the latter two ions merged to form a ``mixture'', i.e. the two ions started to circulate together, which is a well known property of ions in the ESR that have very similar revolution frequencies \cite{Li05}. During such periods, the individual ion frequencies cannot be determined. This effect results in a significant loss of corresponding events from Fig.~2, where counts were only recorded when the ions were clearly separated for at least 20 s.

The second isomer, with $t_{1/2} = 7.0(3)$ min, has had a tentative $I\geq 16$ spin assignment with an excitation energy $>$1910 keV \cite{Br05}. The latter limit assumes log$ft > 5.1$ for allowed $\beta ^-$ decay with 100\% branching to the 2922 keV isomer in $^{212}$Po \cite{Br05}, which itself $\alpha$ decays. Until now, the $^{212m2}$Bi isomer had only been detected through its $\beta$-delayed $\alpha$ decay. The present $^{212m2}$Bi excitation-energy measurement of 1478(30) keV is in clear disagreement. 

Before discussing the interpretation of the new $^{212}$Bi data, it is appropriate to comment on some features of Fig.~2, where ground states and isomers are identified corresponding to lead, bismuth and polonium $A=212$ isobars in the 81+ charge state. First, note that the ground state of $^{212}$Po would be off the scale (at higher frequency) but it is anyway absent due to its short (0.3 $\mu$s) half-life \cite{Br05}; and while the 2922 keV, $t_{1/2}=45$-s $^{212}$Po isomer may be present, it would be unresolved from the $^{212}$Pb ground state. However, it is a special feature of the storage-ring data that, since the $^{212}$Pb ground state is relatively long-lived ($t_{1/2} = 10.6$ h \cite{Br05}), it is possible to distinguish with a good degree of accuracy between $^{212m}$Po and $^{212g}$Pb on an ion-by-ion basis. Thus, of the fourteen 81+ ions observed with the appropriate revolution frequency, ten were assigned to $^{212g}$Pb and four were assigned to $^{212m}$Po. 

As presented later, the peak in Fig.~2 at 125.32 kHz is well explained, by comparison with shell-model calculations, as a high-spin isomer in $^{212}$Bi. This receives strong support experimentally, because the peak is too intense to be an isomer in either $^{212}$Pb or $^{212}$Po, leaving only the $^{212}$Bi possibility. It is nevertheless remarkable that the first isomer of odd-odd $^{212}$Bi is more strongly populated than the $^{212}$Bi ground state. This is due to the isomer's low excitation energy and high spin, enabling it to form an yrast trap. The second isomer is also strongly populated and must be another yrast trap. The population can be quantified in terms of 
the isomeric ratio, which is defined as the ratio of the number of ions of a given nuclide produced in an isomeric state to the total number of ions of that nuclide.
Allowing for merged ions, discussed above, the isomeric ratio is measured to be 63\% for the first isomer of $^{212}$Bi, and 24\% for the second isomer. Only 13\% of the production of $^{212}$Bi goes directly to the ground state.

We propose that the different energies for the second isomer, $>$1910 keV from log$ft$ considerations \cite{Br05} and 1478(30) keV now measured, can be understood through the existence of strong, though unobserved, internal transitions (IT), i.e. $\gamma$-ray or conversion-electron emission. In that case, the {\it partial} $\beta$-decay half-life would be greater than 7 min, so that the previously determined limit on the excitation energy would be correspondingly reduced.

The above proposition has two specific consequences. One concerns the measurement of isomeric ratios. With a large IT branch, the previous restriction to $\beta$-delayed $\alpha$-particle detection \cite{Ba78,Es84} would have led to underestimation of the relative population of the isomer. Indeed, the surprisingly low isomer population following $^{18}$O on $^{208}$Pb reactions was discussed by Eskola et al.~\cite{Es84}. Approximately equal population of the two $^{212}$Bi isomers was expected from ``sum rule model'' \cite{Wi82} calculations, but a maximum cross-section ratio of $\sigma _{m2}/\sigma _{m1} = 0.04$ was observed \cite{Es84}, and this remained unexplained. In contrast, the high isomer population strength seen in the present work (discussed above) involves no assumption about the decay modes. These apparent population differences can be reconciled, at least qualitatively, if there is substantial (unobserved) IT decay, giving a good overall understanding.

The second consequence of the proposed IT branch from the isomer is that a longer half-life should be manifest in the ESR. This is because the measured 7-min half-life of the neutral-atom isomer would be most likely associated with one or more low-energy IT decays that have high electron-conversion coefficients \cite{Ki08}. With only a small number of bound atomic electrons, the highly charged ions in the ESR are subject to suppressed internal conversion \cite{Li03} and hence they can have longer half-lives for nuclear decays. In total, 44 ions of the second isomer of $^{212}$Bi were observed in the present work, having charge states of 80+, 81+ and 82+, i.e. three-, two- and one-electron ions, respectively (while the fully stripped, 83+ ions were out the the ESR acceptance range). However, only three of these ions were lost from the ESR during their 4-min observation periods, due to either $\beta$ decay, electron conversion, or atomic-electron stripping or recombination, and the cumulated isomer-observation time was 182 min. From this, the stored ions are calculated (after Lorentz correction) to have a half-life of 30 min. Since some of the losses could be due to atomic-electron stripping or recombination, the nuclear ($\beta$ plus IT) half-life must be at least 30 min. This is substantially greater than the neutral-atom value of 7 min, which supports the initial proposal that there is a significant IT component. Indeed, a neutral-atom IT branch of $\geq$75\% is implied. Furthermore, the newly measured excitation energy of 1483 keV combined with $t_{1/2} \geq 30$ min implies log$ft \geq 5.1$, fulfilling the previously applied limit \cite{Br05} for allowed $\beta ^-$ decay, i.e. a consistent interpretation is obtained.

Shell-model calculations of the excited states of $^{212}$Bi were carried out with the OXBASH code \cite{Br04} by Warburton \cite{Wa91}, who considered in detail the possible isomer spin and parity assignments. The ground state and first isomer were interpreted as having $I^\pi = 1^-$, and $8^-$ or $9^-$, respectively. The first isomer had a calculated excitation energy of 303 keV ($8^-$) or 281 keV ($9^-$), both of which can be considered to be in satisfactory agreement with the experimental energy of 250(30) keV \cite{Au03}, or 239(30) keV from the present work. The second isomer was interpreted to be an $I^\pi = 18^-$ state, calculated at 1496 keV. This is now observed to be at 1478(30) keV, in excellent agreement.

Due to the need to have more detailed wave-function information (see later) the shell-model calculations were repeated with the same interactions. This revealed a shift of the relative ground-state energy by 40 keV, such that all excited states (other than 1$^-$ states) have energies that are lower by 40 keV, i.e. the 18$^-$ isomer is now calculated to lie at 1456 keV. The reason for the discrepancy remains unexplained. Nevertheless, the 40 keV shift is not large, and the newly calculated 18$^-$ energy of 1456 keV is still in excellent agreement with our measured value of 1478(30) keV. For the first isomer, there is a small improvement in the energy comparisons between calculation and observation -- see Table I for a summary of the values, and Fig.~3 for a partial level scheme.

As discussed by Warburton \cite{Wa91}, the second isomer can be associated with the calculated $I^\pi = 18^-$ state, now at 1456 keV, which is in agreement with the present measurement of 1478(30) keV, but inconsistent with the previous value of $>$1910 keV \cite{Br05}. According to the calculations, the highest-spin state at lower energy than the isomer is an $I^\pi = 15^-$ state 35 keV lower (see Fig.~3). In that case, a 35 keV, $M3$ decay from the isomer would be possible, with an energy that is less than the bismuth $K$-binding energy of 91 keV, with a (neutral atom) conversion coefficient of $\alpha = 3\times 10^5$ \cite{Ki08}, and with a single-particle neutral-atom half-life of 3 s. Compared to the measured neutral-atom half-life of 7 min (previously associated with $\beta$ decay) this would imply a Weisskopf hindrance factor of 140, which is a reasonable value \cite{Fi96} and is in qualitative accord with the present suggestion of a large IT branch (though the calculated energies are not sufficiently precise to come to a definite conclusion). A summary of the experimental and calculated isomer energies is given in Table I. The maximally aligned $\pi h_{9/2}, \nu i_{11/2} (g_{9/2})^2$ configuration for the 18$^-$ state is calculated to have 98\% purity.

Returning to the issue of induced isomer depopulation, $^{212m2}$Bi has some special features. In contrast to the high-multipole ($\lambda = 3$) IT decay, there is the additional possibility to excite states above the isomer by $\lambda = 2$ and $\lambda = 1$ transitions. States with $I^\pi = 16^-$ and $17^-$ are calculated to lie at 1541 and 1628 keV, respectively (see Fig.~3) and these have substantially the same structure (94\% and 95\% purity, respectively) as the 1456-keV $18^-$ isomer, though clearly the orbitals are no longer maximally aligned. Therefore, there is a realistic possibility to induce 85 keV, $E2$ and 172 keV, $M1/E2$ transitions. Using proton and neutron effective charges of 1.5e and 0.5e, respectively, for the $E2$ transition, an excitation strength of 0.7 Weisskopf units is calculated.
Such excitations could be strongly enhanced by the NEEC process in a suitable environment of highly charged ions \cite{Pa07,Go07}, and subsequent de-excitation would be able to take place internally to lower-spin states by conversion-electron and $\gamma$-ray emission, bypassing the isomer. These decay transitions could then give a signal that induced de-excitation had indeed taken place. This system may thus provide a test case for further study, with the large isomeric ratio being an important feature.

In summary, ground-state and isomer observations of $A=212$ isobars in the ESR have provided significant new information on the energy, half-life and population strength of $^{212m2}$Bi. While only $\beta$ decay from this second isomer had previously been identified, the new data give the first direct observations of the isomer and indicate that IT decay competes strongly with $\beta$ decay. It would clearly be desirable to identify experimentally the IT decay radiations. There is also the possibility to search for induced isomer de-excitation.

We thank the technical staff of the accelerators, the FRS, and the target laboratory for their valuable contributions to the beam quality and experimental setups. We also thank the Helmholtz Association of German Research Centers (VH-NG-033), and the UK STFC and AWE plc. for their support.

\clearpage

\begin{table}
\caption{$^{212}$Bi isomers studied in the ESR.}
\vspace{0.2cm}
\begin{tabular}{c|c|c|c|c|c}
              & $I^\pi _{calc}$   & $E_{calc}^{\,a)}$ & $E_{calc}^{new}$  & $E_{exp}^{ESR}$   & $E_{exp}^{\,b)}$  \\
 &  & (keV) & (keV) & (keV) & (keV) \\
\hline
$m1$    & $8^-$, $9^-$   &  303, 281    &  263, 241  & 239(30)  &  250(30)      \\
$m2$    & $18^-$  & 1496    &  1456 & 1478(30) & $>$1910       \\
\hline
\end{tabular}
\end{table}
$^{a)}$ Calculated by Warburton \cite{Wa91}. 

$^{b)}$ Literature excitation energies \cite{Au03,Br05}.

\clearpage

\begin{figure}[t!]
\begin{center}
\includegraphics[width=0.45\textwidth]{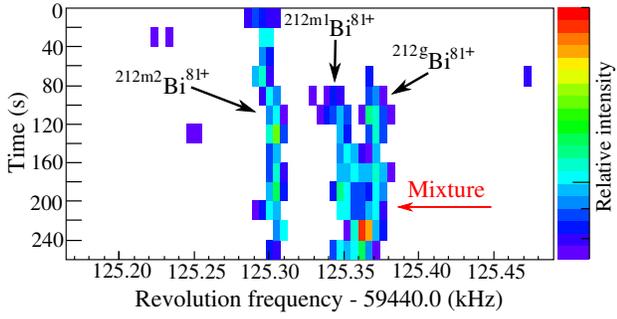}
\caption{(Color online) $^{212}$Bi data, illustrating the ground and two isomeric states as a function of time. The revolution frequency is inversely proportional to the mass-to-charge ratio. During the second half of the observation period, the ground-state and first-isomer ions merge (see text).} 
\end{center}
\end{figure}

%\clearpage

\begin{figure}[t!]
\begin{center}
\includegraphics[width=0.45\textwidth]{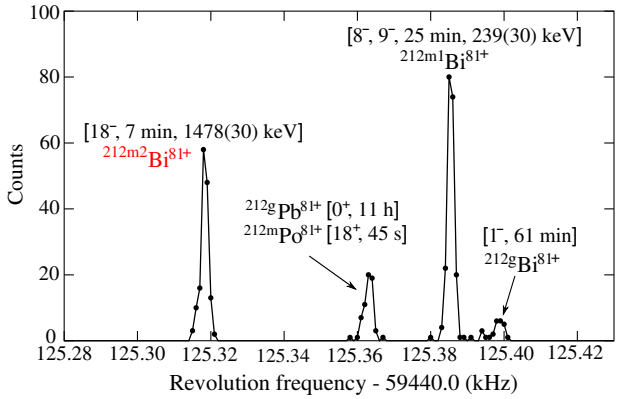}
\caption{(Color online) $^{212}$Bi events, illustrating the ground and two isomeric states, together with unresolved events for $^{212}$Pb and $^{212m}$Po. Each count corresponds to 20 s of observation time. Note that the revolution frequency has had 59440 kHz subtracted. The peaks are labeled with spin/parity values and neutral-atom half-lives, and excitation energies from the present work are included for the $^{212}$Bi isomers.} 
\end{center}
\end{figure}

%\clearpage

\begin{figure}[t!]
\begin{center}
\includegraphics[angle=0,width=0.3\textwidth]{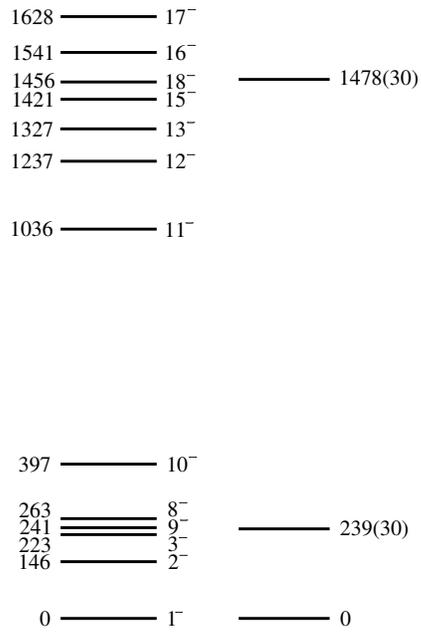}
\caption{Partial level scheme for $^{212}$Bi, showing the calculated energies of the yrast states on the left, together with a few non-yrast states (8$^-$, 16$^-$ and 17$^-$) that are discussed in the text. On the right are the observed isomers with their energies measured in the present work.} 
\end{center}
\end{figure}

%\clearpage

\end{document}